\documentstyle[preprint,eqsecnum,aps]{revtex}
\newif\iftightenlines\tightenlinesfalse
\tightenlines\tightenlinestrue
\begin{document}
%
\def\pT{p_T^{\phantom{7}}}
\def\MW{M_W^{\phantom{7}}}
\def\ET{E_T^{\phantom{7}}}
\def\bh{\bar h}
\def\lm{\,{\rm lm}}
\def\lo{\lambda_1}
\def\lt{\lambda_2}
\def\pslt{p\llap/_T}
\def\eslt{E\llap/_T}
\def\eslt{E\llap/_T}
\def\to{\rightarrow}
\def\Re{{\cal R \mskip-4mu \lower.1ex \hbox{\it e}}\,}
\def\Im{{\cal I \mskip-5mu \lower.1ex \hbox{\it m}}\,}
\def\SU{SU(2)$\times$U(1)$_Y$}
\def\te{\tilde e}
\def\tl{\tilde l}
\def\tb{\tilde b}
\def\tst{\tilde t}
\def\ttau{\tilde \tau}
\def\tg{\tilde g}
\def\tga{\tilde \gamma}
\def\tnu{\tilde\nu}
\def\tell{\tilde\ell}
\def\tq{\tilde q}
\def\tw{\widetilde W}
\def\tz{\widetilde Z}
\def\cmsec{{\rm cm^{-2}s^{-1}}}
\def\fb{{\rm fb}}
\def\sgn{\mathop{\rm sgn}}

\hyphenation{mssm}
\def\ds{\displaystyle}
\def\ts{${\strut\atop\strut}$}
%
%
\preprint{\vbox{\baselineskip=14pt%
   \rightline{FSU-HEP-950204}\break
   \rightline{UH-511-817-95}
}}
\title{SIGNALS FOR MINIMAL SUPERGRAVITY\\
AT THE CERN LARGE HADRON COLLIDER:\\
MULTI-JET PLUS MISSING ENERGY CHANNEL}
\author{Howard Baer$^1$, Chih-hao Chen$^1$, Frank Paige$^2$\\
 and Xerxes Tata$^3$}
\address{
$^1$Department of Physics,
Florida State University,
Tallahassee, FL 32306 USA
}
\address{
$^2$Brookhaven National Laboratory,
Upton, NY 11973 USA
}
\address{
$^3$Department of Physics and Astronomy,
University of Hawaii,
Honolulu, HI 96822 USA
}
\date{\today}
\maketitle
\begin{abstract}

We use ISAJET to perform a detailed study of the missing transverse
energy $\eslt$ plus multi-jet signal expected from superparticle
production at the CERN LHC.  Our analysis is performed within the
framework of the minimal supergravity model with gauge coupling
unification and radiative electroweak symmetry breaking.  We delineate
the region of parameter space where the $\eslt$ supersymmetry signal
should be observable at the LHC and compare it to the regions
explorable via searches for sleptons and for chargino/neutralino
production.  We confirm that, given a data sample of 10~$\fb^{-1}$,
$m_{\tg}\sim 1300$ GeV can be explored if $m_{\tq}\gg m_{\tg}$, while
$m_{\tg}\sim 2000$ GeV can be probed if $m_{\tq}\simeq m_{\tg}$.  We
further examine what information can be gleaned from scrutinizing this
event sample. For instance, the multi-jet multiplicity yields
information on whether squark production makes a significant
contribution to the observed $\eslt$ sample. Furthermore,
reconstructing hemispheric masses may yield a measure of $m_{\tg}$ to
$\sim 15-25\%$.  Finally, for favourable ranges of parameters, by
reconstructing masses of tagged $b\bar{b}$ jet pairs, it may be
possible to detect Higgs bosons produced via sparticle cascade decay
chains.

\end{abstract}

\medskip



\section{Introduction}

\subsection{Motivation and Framework}

The existence of weak-scale supersymmetry (SUSY), stabilizing the
electroweak symmetry breaking sector of the Standard Model (SM), is a
tantalizing hypothesis\cite{REV}.  This hypothesis has received some
support from the observation that the running gauge couplings unify at
a scale $M_X\sim 10^{16}$~GeV provided that the superpartners have
masses $\sim 1$~TeV, while this unification does not occur in the
SM\cite{UNIF}.  Hence, the search for supersymmetric particles is a
high priority task for collider experiments.  Expectations for
super-particle masses are typically in the 100--1000~GeV range,
whereas current limits on them are generally below
$M_Z/2$\cite{LEPREV}, although the CDF and D0 experiments at the
Tevatron probe gluinos and squarks as heavy as
150--200~GeV\cite{TEVGL}.  In the near future, LEP~II will probe
slepton and chargino masses up to $m_{\tl},m_{\tw_1}\sim 90$ GeV,
while Tevatron experiments, for favourable parameter ranges, may
indirectly reach $m_{\tg}\sim 500$ GeV in the main injector era,
via the clean tri-lepton signal from $\tw_1\tz_2$
production\cite{TEVSTAR,KAMON,DPF}.  These experiments have a good
chance to discover weak scale supersymmetry, but they cannot exclude it.  A
thorough search for low-energy supersymmetry can only be made at the
recently approved CERN Large Hadron Collider (LHC) or, alternatively,
at $e^+e^-$ linear colliders with center of mass energies exceeding
500-1000~GeV\cite{JLC,DPF}.

Most early analyses were carried out
within the framework of the Minimal Supersymmetric Model
(MSSM), which is the supersymmetric extension of the SM with the least
number of additional new particles and interactions necessary for
phenomenology. As a consequence, $R$-parity is exactly conserved, so
that the lightest SUSY particle (LSP) is stable.  Supersymmetry
breaking is parametrized by the introduction of soft SUSY breaking
mass terms and interactions consistent with the SM gauge group. Since
each $SU(3)\times SU(2)\times U(1)$ multiplet of sfermions and
gauginos has an independent mass, the most general parametrization of
SUSY breaking requires a large number of mass parameters, and also
many arbitrary trilinear scalar couplings.

Without further assumptions about these parameters, any
phenomenological analysis is essentially intractable. Motivated by
grand unified theories (GUTs),
we shall, following the early SUSY analyses in Ref.\cite{EARLY,sdc,gem},
further assume that the three $\overline{DR}$ gaugino
masses originate in a common gaugino mass at some high scale, so that
the ratios of these masses at the weak scale are proportional to the
corresponding ratios of the squared gauge couplings. The $SU(2)$ and
$U(1)$ gaugino masses are then determined in terms of $m_{\tg}$.
Motivated by supergravity models, these
analyses further assumed a  common {\it physical}
mass for each of the light squarks, left and
right sleptons, and sneutrinos.
The Higgs sector of this model is determined at
tree level
by $m_{H_p}$, the mass of
the neutral pseudoscalar, in addition to
the parameters $\tan\beta$, the ratio of the two vacuum
expectation values, and $\mu$, the superpotential Higgsino mass (which also
enter the gaugino-Higgsino sector).
Finally, the weak-scale $A$-parameters,
which mainly affect the phenomenology of third generation squarks were
needed to completely specify the model. In
what follows, and in the simulation program ISAJET\cite{ISAJET},
we use the term MSSM to refer to  this
supergravity-inspired framework.

In view of the fact that the
additional assumptions for the MSSM are motivated by
supergravity GUT models, it seems reasonable to seriously explore their
implications. The assumptions underlying these models, which
differ in important ways from those of the MSSM, are outlined below.
Several
phenomenological analyses of supergravity models\cite{FN1} have recently
appeared in the literature\cite{MUR,LOPEZ,BDKNT,BCMPT}.  In minimal
supergravity with canonical kinetic energy terms, gravitational
interactions communicate the effects of supersymmetry breaking in a
hidden sector to the observable sector of particles.  These
gravitational interactions, being universal, lead to a common mass
($m_0$) for all scalar fields and common trilinear and bilinear soft
SUSY breaking ($A_0$ and $B_0$) scalar interactions\cite{SUGRAV}. The
resulting Lagrangian should be regarded as an effective Lagrangian of
the theory below the Planck scale, with its parameters renormalized at
some ultra-high energy scale close to $M_{Planck}$. If we further
assume that these interactions respect an (unbroken) GUT symmetry, the
gauginos must all be part of a single multiplet, and so must have a
common mass ($m_{1/2}$) at the this scale. Similar considerations also
apply to string models, which can lead to coupling constant
unification without a grand unifying gauge group\cite{STRING}. The
SUSY breaking sector of the model is then completely specified by
these four parameters.

For phenomenological analyses, the running parameters in the
Lagrangian should be renormalized at the weak scale to sum the large
logarithms arising from the disparity between the high scale where the
physics is simple, and the weak scale relevant to collider
phenomenology. This is most conveniently done using renormalization
group (RG) equations\cite{RGE} and taking the common values of the four
high-scale input parameters as boundary conditions.  The RG evolution
splits the various masses and leads to a rich spectrum at the weak
scale\cite{SUGRA}.  The first two generations of squarks are
approximately degenerate, so these models automatically satisfy
constraints\cite{FCNC} from the non-observation of flavour changing
neutral currents in the kaon sector, while $\tst_1$, the lighter
$t$-squark and $\tb_1 \sim\tb_L$, the lighter $b$-squark, may be
substantially lighter.  Sleptons are lighter than squarks, and may be
much lighter if gluinos and squarks of the first two generations have
comparable masses.  As a bonus, the RG evolution also gives the
correct pattern of electroweak symmetry breaking\cite{RAD,RGE} for
considerable ranges of input parameters.  It is customary to eliminate
$B_0$ in favour of $\tan\beta$ and to adjust the value of $\mu$ to
reproduce the measured $Z$ boson mass.  Then the particle/sparticle
masses and couplings are determined by the following parameters (along
with $m_t$):
\begin{eqnarray}
m_{0},\ m_{1/2},\ A_0,\ \tan\beta,\ {\rm and}\ \sgn(\mu ).
\eqnum{1}
\end{eqnarray}
In particular, the values of the MSSM parameters $\mu$ and $m_{H_p}$
are determined. Typically, $\mu$ is large, so that the lighter
charginos and neutralinos are dominately gaugino-like, while the
heavier ones are dominantly Higgsino-like. Also, $m_{H_p} \gg M_W$ so
the lightest Higgs scalar ($H_{\ell}$) resembles a SM Higgs boson with
the other Higgs particles all relatively heavy. In what follows, we
shall use the term SUGRA to mean supersymmetry with masses and mixings
calculated from the parameters in equation (1).

The simulation program ISAJET\cite{ISAJET} now allows the user either
to specify the weak-scale MSSM parameters --- the gluino, squark,
stop, sbottom and slepton masses, $A$-parameters, $\mu$, $m_{H_p}$ and
$\tan\beta$ --- or to use values of these parameters calculated from
the SUGRA parameter set (1).  Because the SUGRA model is specified by
just four new parameters, various experimental observables that can be
determined in experiments at colliders become correlated and so
provide non-trivial tests of the underlying assumptions\cite{MUR,BCMPT,DPF}.
It is, however, worth remembering that SUGRA models are based on
extrapolations of physics which may ultimately prove incorrect. While
these models are indeed very attractive, economical, and satisfy all
phenomenological constraints, it may be worthwhile to test the
sensitivity of model predictions by running ISAJET using the optional
MSSM input set, which is a more general framework that can encompass
models with non-universal soft-breaking sfermion mass terms.  This may
be particularly important for assessing the reach of future facilities.

\subsection{Phenomenological Overview}

Since strongly interacting sparticles are most copiously produced at
hadron colliders, many of the early studies of supersymmetry at hadron
supercolliders focussed on the detection of gluinos and
squarks\cite{EARLY,sdc,gem}.  Recently, a number of papers addressing
the detection of weakly interacting sparticles at the LHC have also
appeared.  In Ref.~\cite{slep}, it was shown that LHC experiments
ought to be able to detect clean dileptons from slepton pair
production if $m_{\tl}\stackrel{<}{\sim}200$--$250$~GeV.  Also, in
Ref.~\cite{trilep}, it was shown that LHC experiments ought to be able
to detect the clean trilepton signal from $\tw_1\tz_2$ production over
much of parameter space as long as the two body decays $\tz_2 \to
H_{\ell}\tz_1$ and $\tz_2 \to Z\tz_1$ are kinematically inaccessible.
Experiments at a high luminosity upgrade of the Tevatron may have a
similar reach as the LHC if $\mu$ is negative; for positive values
of $\mu$ the branching fraction for the three body decays
$\tz_2\to\ell\bar{\ell}\tz_1$ is strongly suppressed, and the
corresponding reach at the Tevatron is somewhat smaller than at the
LHC.

Direct slepton pair and chargino/neutralino production takes place via
weak interactions, whereas the strength of the LHC lies in the
production of strongly interacting particles. In Fig.~\ref{fig1}, we
show the total production cross sections at $\sqrt s = 14\,{\rm TeV}$
for strongly interacting SUSY particles ($\tg\tg$, $\tg\tq$ and
$\tq\tq$), for charginos and neutralinos in association with squarks
and gluinos, and for gaugino pairs ($\tw_1\tz_2$ and $\tw_1\tw_1$), as
a function of $m_{\tg}$.  We have assumed gaugino mass unification and
5 degenerate species of L- and R-squarks, taken $\mu =-m_{\tg}$ and
$\tan\beta =2$, and used the CTEQ2L parton distribution
functions\cite{CTEQ}. In Fig.~\ref{fig1}{\it a}, we take
$m_{\tq}=m_{\tg}$, while in Fig.~\ref{fig1}{\it b} we take
$m_{\tq}=2m_{\tg}$.  In {\it a}), it is clear that strong sparticle
production is the dominant production mechanism at the LHC for all
values of $m_{\tg}$ from 300 GeV out to 2000 GeV.  In {\it b}), strong
sparticle pair production is dominant up to $m_{\tg}\sim 1100$ GeV,
after which chargino/neutralino production becomes dominant.  In both
cases, associated production mechanisms are sub-dominant cross
sections, even though the combined mass of the produced sparticles can
be much smaller than the mass of a pair of strongly interacting
sparticles.  The fact that strongly-produced sparticle pairs are the
dominant production mechanism for squarks and gluinos as heavy as
1--2~TeV leads us to focus on signals from squarks and gluinos.

Once gluinos and squarks are produced, they are expected to decay
through various channels until the cascade terminates in the stable
LSP (taken to be $\tz_1$)\cite{CASCADE}.  These cascade decays lead to
final states containing multiple jets, isolated and non-isolated
leptons, and missing transverse energy (mainly from the undetected
$\tz_1$'s and neutrinos).  Rates for the various multi-lepton
signatures have been presented in Ref.~\cite{btw}; the most promising
signatures appeared to be those containing same-sign isolated
dileptons\cite{bgh}, and isolated trileptons.  The ${\rm multi-jet} +
\eslt$ signal, while generally much larger than signals from
multi-leptons, is less clean, due to irremovable backgrounds from
various SM QCD induced processes (multi-jet production, vector boson
production in association with jets and heavy flavor production).

Detailed studies of the missing energy plus multi-jet signal have been
performed by the GEM\cite{gem} and SDC\cite{sdc} collaborations for
the now defunct Superconducting Supercollider project\cite{franksim}.
More recently, detailed studies of this same signature have also been
performed by the ATLAS\cite{atlas} Collaboration for the CERN LHC.  In
these LHC studies, performed within the framework of the
supergravity-inspired MSSM, it was shown that values of $m_{\tg}\sim
1300$ GeV ($m_{\tg}\sim 2000$ GeV) could be probed for $m_{\tq}\gg
m_{\tg}$ ($m_{\tq}\sim m_{\tg}$), given 10 $\fb^{-1}$ of integrated
luminosity.  In addition, values of $m_{\tg}$ as low as 300 GeV could
easily be probed at the LHC, so that there should be no ``gap'' in the
explorable range of $m_{\tg}$ between Tevatron searches and future LHC
searches.

In the present paper, we further scrutinize the missing energy plus
multi-jet signal. Our goals are multiple:
\begin{enumerate}
\item We update previous results\cite{btw} on multijet $+\eslt$ cross
sections by using the latest ISAJET 7.13 simulation for sparticle
production and cascade decays.
\item We evaluate the reach of the LHC via some set of optimized cuts,
valid across a large range of sparticle mass choices.
\item We present our reach results in the parameter space of the
minimal supergravity model with gauge coupling unification and
radiative electroweak symmetry breaking. The rather small parameter set
yields a correlated spectrum of SUSY particle masses, and allows one to
compare the regions of parameter space that different search
experiments can probe.
\item We examine what information can be gleaned from a sample of
signal events in the missing energy plus multi-jet channel. For
instance, can one tell whether the signal is mainly due to gluino
production, or whether $\tq\tq$ and $\tq\tg$ production is also
significant?  Can one gain some sort of mass measurement for the
gluino or squark? Can one identify the presence of other particles or
sparticles by sifting through the debris of the cascade decay?
\end{enumerate}

The first three of these points are addressed in Sec.~II of this paper,
where we mainly map out regions of SUGRA parameter space explorable by
LHC experiments using the multi-jet plus missing energy signature. We
also compare these regions with those where various slepton and
chargino/neutralino searches are expected to result in observable
signals.  In Sec.~III, we address the questions raised above in
item~(4), and find that to some degree, and at least in some cases,
all the questions raised can be answered in the affirmative. We
summarize our results and present our conclusions in Sec.~IV.

\section{Reach of the LHC in SUGRA parameter space}

\subsection{Event simulation}

In this paper, we work within the framework of the minimal SUGRA model,
as implemented in the ISAJET 7.13 subprogram ISASUGRA\cite{BCMPT}.
The SUGRA parameter set (1) can be directly entered into ISAJET, and
ISASUGRA then calculates all the SUSY particle masses and mixings.
Briefly, starting with the precisely measured gauge couplings at scale
$M_Z$, ISASUGRA evolves the three MSSM gauge couplings and third
generation Yukawa couplings to a mass scale (determined to be $\sim
2\times 10^{16}$ GeV) where $\alpha_1$ and $\alpha_2$ unify. At this
approximate unification scale, $\alpha_3$ is set equal to $\alpha_1$
and $\alpha_2$, and the GUT scale boundary condition values for $m_0$,
$m_{1/2}$, $A_0$ and $B_0$ are implemented. From the approximate
unification scale, the various soft SUSY breaking masses, gauge and
Yukawa couplings are evolved via 26 RGE's to their weak scale values.
We use 2-loop RGE's for gauge couplings, but only 1-loop RGE's for
soft breaking terms. Weak scale sparticle threshold effects are
included within the gauge coupling evolution. Soft breaking masses are
evolved only down to the scale value equal to their mass value. At the
electroweak scale, the 1-loop effective potential is minimized,
allowing the replacement of $B$ by $\tan\beta$, and evaluating the
magnitude (but not the sign) of $\mu$ in terms of $M_Z$. This
procedure is iterated until a stable solution for all sparticle masses
is obtained, consistent with grand unification and electroweak
symmetry breaking.

After calculation of the spectra and couplings as detailed above,
ISASUGRA then calculates all available sparticle decay modes and
branching fractions. Currently, the program is valid only for
$\tan\beta\alt 10$; for very large values of $\tan\beta$, the effects
from sbottom and stau mixing, not yet included, become important.  It
also assumes that the LSP is $\tz_1$, which is not true for all
choices of the parameters in set~(1). Next, ISAJET generates all
lowest order $2\to 2$ subprocesses for sparticle pair production
according to their cross sections. The produced sparticles are then
allowed to decay via the various possible cascades. Initial and final
state parton showers are implemented, as well as hadronization and
beam fragment evolution.

For detector simulation at the LHC, we use the toy calorimeter
simulation package ISA\-PLT.  We simulate calorimetry covering
$-5<\eta <5$ with cell size $\Delta\eta\times\Delta\phi =0.05\times
0.05$. We take the hadronic energy resolution to be $50\%
/\sqrt{E}\oplus 0.03$ for $|\eta |<3$, where $\oplus$ denotes addition
in quadrature, and to be $100\% /\sqrt{E}\oplus 0.07$ for $3<|\eta
|<5$, to model the effective $p_T$ resolution of the forward
calorimeter including the effects of shower spreading.  We take
electromagnetic resolution to be $10\% /\sqrt{E}\oplus 0.01$.  Jets
are found using fixed cones of size $R=\sqrt{\Delta\eta^2
+\Delta\phi^2} =0.7$ using the ISAJET routine GETJET.  Clusters with
$E_T>100$ GeV and $|\eta ({\rm jet})|<3$ are labeled as jets.  Muons
and electrons are classified as isolated if they have $p_T>20$ GeV,
$|\eta (\ell )|<2.5$, and the visible activity within a cone of $R
=0.3$ about the lepton direction is less than $E_T({\rm cone})=5$ GeV.

\subsection{Signal versus background}

To evaluate signals from supersymmetry in the multi-jet $+\eslt$
channel, we generate all possible supersymmetric subprocesses using
ISAJET. Major physics backgrounds come from various SM processes which
can give large amounts of missing transverse energy due to neutrinos
produced in events, due to mis-measurement by calorimeter cells, and
due to dead regions of the detector. We evaluate to following SM
background processes:
\begin{itemize}
\item QCD multi-jet production ({\it e.g.} $gg\to gg$ {\it etc.}, where
extra jet activity comes from parton showers, (this includes heavy flavor
$b\bar b$ and $c\bar c$ production),
\item $Z+$multi-jet production, where $Z\to\nu\bar{\nu}$ or $\tau\bar{\tau}$,
\item $W+$multi-jet production, where $W\to\ell\nu_{\ell}$ or $\tau\nu_{\tau}$,
\item $t\bar t$ production and decay.
\end{itemize}

We impose a series of cuts to extract signal from the vastly larger SM
production cross sections. Since we wish to detect gluinos or squarks
over a large mass range $\sim 300$--$2000$~GeV, different cuts are needed
to optimize signal/background depending on the sparticle mass. After
exploring a large variety of possible cuts for various SUGRA parameter
choices, we arrived at the following requirements:
\begin{itemize}
\item no isolated leptons,
\item transverse sphericity $S_T>0.2$ to reduce QCD dijet background,
\item number of jets $n(jets)\geq 2$ (jets as defined above),
\item transverse plane angle $\Delta\phi (\vec{\eslt},j_c )$
between $\vec{\eslt}$ and closest jet is $30^o<\Delta\phi <90^o$.
\end{itemize}
After these mass-independent cuts, we apply a variable cut with the
parameter $E_T^c$ chosen depending on the gluino and squark masses:
\begin{itemize}
\item $\eslt >E_T^c$\ {\rm and}\ $E_T(j_1),\ E_T(j_2)>E_T^c$.
\end{itemize}

The results with these cuts for signal and total background levels are
shown in Fig.~\ref{fig2}, versus the cut parameter $E_T^c$. We show
the signal cross sections for the six cases listed in Table~I, which
roughly correspond to $m_{\tg}\sim 300,\ 800$ and $1300$ GeV, where
$m_{\tq}\sim m_{\tg}$ or $m_{\tq}=(1.5-1.7)m_{\tg}$. We also show the
estimated background from the sum of all SM processes. Since the total
SM cross section corresponds to about $10^{15}$ events, it is
obviously neither technically possible nor physically reasonable to
generate realistic statistics for it.  Instead, the various SM
processes have been generated in several overlapping $p_T$ ranges to
obtain a reasonable estimate for all $E_T^c$. The background is
discussed in more detail in Sec.~II{\it c} below.

 From Fig.~\ref{fig2}, we see that for $E_T^c=100$ GeV, case 1 and 2
with $m_{\tg}\sim 300$ GeV are easily visible at levels of $5$--$10$
above background, while the cases with heavier gluino masses are well
below background. As the cut $E_T^c$ is increased, the SM background
decreases much more quickly than the heavy gluino signals. For
$E_T^c>150$--$200$~GeV, the cases with $m_{\tg}\sim 800$ GeV begin
exceeding background. To see the signal from $m_{\tg}\sim 1300$ GeV
above background, a hard cut of $E_T^c>300$--$400$~GeV is needed.  This
agrees qualitatively with the results of Ref.~\cite{atlas}, where two
sets of cuts (soft and hard) were advocated to see light or heavy
gluinos. As we will see shortly, gluino and squark mass values across
the range of $m_{\tg}\sim 300$ up to $m_{\tg}\sim 1300$--$2000$~GeV
ought to be detectable. In particular, there should be no
``undetectability window'' in $m_{\tg}$ between Fermilab Tevatron
experiments of the Main Injector era, and LHC experiments

Fig.~\ref{fig2} also suggests a way to get a crude estimate of the
gluino or squark mass.  One plots the observed cross section versus
$E_T^c$ (for gradual stiffening of cuts), and compares with background
expectations. The approximate range where $m_{\tg}$ might lie can be
obtained by measuring event rates for several values of $E_T^c$,
starting at the point where observation begins exceeding expectations
from SM processes, and comparing against Monte Carlo expectation of
the signal.

In Fig.~\ref{fig2}, we have fixed $A_0=0$ and $\tan\beta =2$.
Variations in $A_0$ mainly affect third generation squark and slepton
masses, so our results are relatively insensitive to different $A_0$
values as long as the light stop $\tst_1$ is not driven to too small a
value.  As an example, we plot in Fig.~\ref{fig3}{\it a} the cross
section after cuts (taking $E_T^c=300$ GeV) for $m_0=m_{1/2}=300$ GeV
(case 3) and also for $m_0=2m_{1/2}=600$ GeV, for five choices of
$A_0$. Little variation is seen when $A_0$ varies over the range
shown.  Multi-jet$+\eslt$ signals are also relatively insensitive to
variations in $\tan\beta$.  Case~3 is also plotted in
Fig.~\ref{fig3}{\it b} for five choices of $\tan\beta$ and again shows
only small differences in signal cross section.

\subsection{Reach of the LHC in SUGRA parameter space}

Our next goal is to evaluate the reach of the LHC via the ${\rm
multi-jet} + \eslt$ signature in SUGRA parameter space. From
Fig.~\ref{fig2}, it is clear that for very heavy gluinos, a large
value of $E_T^c$ is desirable to enhance the signal relative to the
background.  For the heaviest gluinos and squarks observable at the
LHC, which are roughly the heaviest consistent with SUSY relevant to
electroweak symmetry breaking, we want to make a cut at about
$E_T^c=500$~GeV.  Only a very small fraction of the SM cross section
passes such a cut, so it is necessary to combine events generated in
several ranges of hard scattering $p_T$ ($p_T^{HS}$)
for each SM process that can produce
backgrounds.  Table~II lists the ranges generated for each process,
the total number of events for each range, and the number of Monte
Carlo events and the corresponding cross section passing the cut
$E_T^c = 500\,{\rm GeV}$.  Fig.~\ref{figbkg} shows the background
cross sections for each process and range of $p_T^{HS}$ vs.\ $E_T^c$ for
those values of $E_T^c$ for which it is non-zero for the available
statistics. While in retrospect a more uniform distribution of Monte
Carlo events in $\log p_T$ would have given a better estimate of the
background, the samples listed in Table~II are sufficient to provide
an estimate of the SM background for all relevant values of $E_T^c$.
In particular, for each $E_T^c$ there are enough events in some range
for each process to give an estimated background cross section greater
than that from lower $p_T$ ranges that produce no events.  Therefore,
we estimate the SM background as a function of $E_T^c$ using the 95\%
upper limit calculated from the ranges giving a nonzero number of
events, setting the background from the ranges giving zero events to
zero.  In particular, the total background from Table~II then comes
out to $1.86\pm 0.36$ $\fb$, which gives a 95\% CL upper bound on the
total BG to be $\sigma = 2.44\ \fb$. We conservatively use this value
to obtain our $5\sigma$ reach.

Since ISAJET generates higher order QCD processes using the branching
approximation, it does not necessarily give the correct background for
SUSY signatures, which typically involve ``round'' events with jets
far from the collinear limit in which the branching approximation is
correct. Fortunately, we find that our estimates of the SM background
is smaller than the SUSY signal, so the SUSY mass reach is not
very sensitive to the background estimate. Also, the significance of
any signal is not dependent on the calculation of the SM backgrounds,
since these can be determined from other data.  Backgrounds from
neutrinos from $b$ and $c$ decays can be checked against the $p_T$
distribution of non-isolated muons.  Backgrounds from $Z \to \nu
\bar\nu$ can be checked against measurements of $Z \to \ell^+\ell^-$.
Backgrounds from $W$ and $t \bar t$ can be checked against
distributions of isolated single and double leptons. Detector induced
backgrounds can be checked using jet-jet and $\gamma$-jet events.
Given this data, it should be possible to determine the SM background,
including the multijet and $S_T$ cuts, accurately and so assign a
significance limited only by statistics to any excess from SUSY.

The $5\sigma$ background level calculated as described above is shown
by the horizontal line in Fig.~\ref{fig4}.  (Notice that the signal to
background ratio is very close to unity even for a signal just at this
level so that we do not need to impose any other S/B requirement for
the observability of the signal.)  On the same plot, we show signal
rates versus $m_{\tg}$ for two cases:  {\it i}) squares for
$m_0=m_{1/2}$ ($m_{\tq}\sim m_{\tg}$), and {\it ii}) triangles for
$m_0=4m_{1/2}$ ($m_{\tq}\gg m_{\tg}$). From this plot, we see that for
$m_{\tq}\sim m_{\tg}$, the LHC should be able to probe to $m_{\tg}\sim
2000$ GeV, while for $m_{\tq}> m_{\tg}$, LHC should be able to probe
to nearly $m_{\tg}\sim 1300$ GeV. These results are in remarkably
close agreement with the reach calculated in Ref.~\cite{atlas}, which
used somewhat different MSSM parameter choices and different cuts.

In Ref.~\cite{BCMPT}, it was shown that in the minimal SUGRA model, the
$m_0\ vs.\ m_{1/2}$ plane forms a convenient panorama in which to plot
various constraints. In Figs.~\ref{fig5}--\ref{fig7}, we plot the
reach of the LHC in the multi-jet$+\eslt$ channel, using the $5\sigma$
constraint for 10~fb$^{-1}$ of collider data. For convenience, in
frame {\it a}) we show contours in SUGRA parameter space; in frame
{\it b}) we show corresponding gluino and squark (averaged over the 4
first generation squarks) mass contours. All the figures take $A_0=0$
and $m_t =170$ GeV.  Fig.~\ref{fig5} plots contours for $\tan\beta =2$
with $\mu <0$, while Fig.~\ref{fig6} plots for the same value of
$\tan\beta$, but with $\mu >0$.  Finally, Fig.~\ref{fig7} shows
results for $\tan\beta =10$ and $\mu <0$.

In these figures, the region shaded with bricks is excluded on
theoretical grounds: either the correct electroweak symmetry breaking
pattern is not obtained (or gives the wrong value of $M_Z$), or the
LSP is not the $\tz_1$, but is instead the $\tw_1$, the $\ttau_1$,
$\tell_R$ or $\tnu_L$. In addition, for some values of $A_0$, the light
stop $\tst_1$ can be driven tachyonic, so that electromagnetic and
colour symmetries are spontaneously broken. The hatched regions are
excluded by experiment. The experimentally excluded regions are
formed\cite{BCMPT} by combining bounds from LEP\cite{LEPREV} that
$m_{\tw_1}>47$ GeV, $m_{H_{\ell}}>60$ GeV, and $m_{\tnu}>43$ GeV. In
addition, the latest bound from CDF/D0 on multi-jets$+\eslt$
events\cite{TEVGL} is included.

In Ref.~\cite{slep}, it was shown that LHC ought to be able to probe
sleptons with mass $m_{\tl}\sim 200$--$250$~GeV. The area to the left of
the contour labelled $\tell (200)$ denotes this region in SUGRA parameter
space. Also, it was shown in Ref.~\cite{trilep} that (for $\mu < 0$)
LHC experiments ought to be able to explore much of the region below
the contours labelled $\tz_2\to\tz_1 H_{\ell}$ and $\tz_2\to\tz_1 Z$
(above which the so-called spoiler modes become accessible), by
searching for the clean trilepton signature of $\tw_1\tz_2\to 3\ell$.
This region also corresponds approximately to the maximal reach of
Fermilab Tevatron collider experiments (although large holes of
non-observability exist for some ranges of parameters due to a strong
suppression of the $\tz_2\to\tz_1\ell\bar{\ell}$ branching fraction,
especially for positive values of $\mu$).  Furthermore, LEP II should
be able to explore the regions below the contours marked $H_l (90)$
and $\tw_1 (90)$ where the lightest Higgs boson and the chargino are,
respectively, lighter than 90~GeV.

In each of Figs.~\ref{fig5}{\it a}--\ref{fig7}{\it a}, we see the
$5\sigma$ $\eslt$ reach is maximal for $m_0\sim {1\over 2} m_{1/2}\sim
500$ GeV, and drops to intercept the theoretically excluded (due to
$m_{\ttau_1}<m_{\tz_1}$) region around $m_0\sim 200$ GeV; this fall-off
is due to the fact that as $m_0$ decreases, sleptons become very
light, resulting in the presence of many isolated leptons in the final
state of gluino and squark cascade decays. The isolated lepton veto
thus suppresses somewhat the multi-jet$+\eslt$ cross section as $m_0$
decreases; we may, however, expect that the multilepton cross section
is large in this range of $m_0$.  For not too large values of $m_0$,
squark masses are somewhat lighter than (or comparable in mass to)
gluinos, so that the combined cross section for $\tg\tg$, $\tg\tq$ and
$\tq\tq$ production is very large.  In this region, the maximal reach
of the LHC is obtained: $m_{\tg}\sim 2000$ GeV can be explored.  For
$m_0$ very large compared to $m_{1/2}$, squarks and sleptons are far
heavier than gluinos.  In this case, superparticle pair production is
dominated by just $\tg\tg$ production (for $m_{\tg}<1100$ GeV).
Nevertheless, a mass reach of $m_{\tg}\sim 1300$ GeV is obtained.

Recently, upper limits on sparticle masses have been obtained by
requiring no fine-tuning in minimal SUGRA models\cite{finetun}. These
limits, which are somewhat subjective, suggest $m_{\tg},\ m_{\tq} \alt
700$--800~GeV. Comparison of these values with the calculated reach of
the LHC suggests that LHC can perform a complete scan over the
parameter space of minimal SUGRA models.

\section{Characteristics of the multi-jet $+\eslt$ signal}

If supersymmetry is discovered, it will be important to see what
information can be gleaned from signal events about the properties of
super-particles. We have already mentioned that a rough determination
of $m_{\tg}$ and $m_{\tq}$ can be made by noting the size of the signal
cross section above background for different choices of the cut
parameter $E_T^c$.  In this section, we focus on several examples of
intrinsic event properties, and the information they provide.

\subsection{Jet multiplicity: Detecting squarks in the $\eslt$ sample}

If $m_{\tg}$ is sufficiently smaller than $m_{\tq}$, then gluino pair
production dominates, and each gluino decays typically via a lengthy
cascade resulting in numerous final state jets, {\it e.g.}, $\tg\to
q\bar{q}\tz_i$, $\tg\to q\bar{q}\tw_i$ or $\tg\to t\tst_1$, followed
by further $\tz_i$, $\tw_i$, $t$ and $\tst_1$ decays.  In contrast,
if $m_{\tq}\alt m_{\tg}$, squarks decay through a more abbreviated
cascade, via $\tq_L\to q\tw_i,\ q\tz_i$, and $\tq_R\to q\tz_i$ (Third
generation squarks can, however, decay to charginos via their
couplings to the Higgsino component). For right squarks, $\tq_R\to
q\tz_1$ dominates over large regions of parameter space. Since hard
jets are most likely to come from the primary decay, in
multi-jet$+\eslt$ events, for a fixed value of $m_{\tg}$, one
frequently expects a higher jet multiplicity if $m_{\tq}>m_{\tg}$, and
$\tg\tg$ production dominates, than if $m_{\tq}<m_{\tg}$, and $\tq\tq$
and $\tq\tg$ production dominates.  (An alternative method involving
the study of the charge asymmetry in the same sign dilepton sample was
suggested in Ref.~\cite{btw}.)

This is explicitly illustrated in Fig.~\ref{fig8} where we plot the jet
multiplicity after cuts (for the $E_T^c$ value listed) for each of the
SUGRA cases listed in Table~1.  We have relaxed the jet cut to require
only $p_T({\rm jet})>50$ GeV for this plot, so that we increase the
sensitivity to lower energy jets produced further down the cascade
decay chain. The histograms shown include both signal and background
contributions. In Fig.~\ref{fig8}{\it a}, we show jet multiplicity for
the two $m_{\tg}\sim 300$ GeV cases (with $E_T^c=150$ GeV), where case
1 has $m_{\tq}\sim m_{\tg}$, and case 2 has $m_{\tq} \gg m_{\tg}$. We
see that case 2 is dominated by 4 and 5 jet production, indicative of
$\tg\tg$ production and decay, whereas case 1 has a significantly
lower average jet multiplicity, and produces mainly 3 and 4 jet events
more characteristic of a large component of $\tq\tq$ production. In
Fig.~\ref{fig8}{\it b}, the two cases for $m_{\tg}\sim 800$ GeV again
show a larger jet multiplicity for the case where $m_{\tg}\ll
m_{\tq}$.  Finally, Fig.~\ref{fig8}{\it c} shows the jet multiplicity
for the two cases with $m_{\tg}\sim 1400$ GeV, and again the
$m_{\tg}\ll m_{\tq}$ case has larger jet multiplicity. For case 6, we
see that the jet multiplicity is large for $n({\rm jet})=3$, and then
falls for $n({\rm jet})=4$, and rises again to a maximum for $n({\rm
jet})=5$. This is due to a large background contribution in the
$n({\rm jet})=3$ bin.

In all cases, the $n({\rm jet})$ distribution has a wide range of
smearing.  This is due to multiple hard partons produced at various
stages along the cascade decay chain, but is also due to substantial
initial and final state QCD radiation. Nonetheless, the final
distributions do show that, if we have some idea of the gluino mass,
the measured jet multiplicity can give a handle on whether or not the
event sample contains a significant $\tq$ production component, thus
reflecting the relative values of $m_{\tg}$ and $m_{\tq}$. Although we
have illustrated this only for a few cases, we expect this to be a
generic feature. The multiplicity in gluino events can only be reduced
if the branching fraction for the radiative decay $\tg \to g\tz_1$
becomes very large.

We have also examined several other techniques for detecting the
presence of squarks in the $\eslt$ sample. These are based on the
expectation that heavy squark events (which form only a small fraction
of the total sample) are expected to be more spherical and have larger
jet multiplicities. For the first two cases in Table~I, we have
examined ({\it i}) the transverse sphericity distribution, ({\it ii})
the scatter plot of the transverse sphericity ($S_T$) versus the
``bigness'' $B= |\eslt| + \Sigma |p_T({\rm jet})| $, ({\it iii})
scatter plot of $B$ versus $k_T({\rm max})$, the largest transverse
momentum relative to the sphericity axis and ({\it iv}) the scatter
plot of $B$ versus $n({\rm jet})$. Only the last of these
distributions appear to show any significant difference. Because this
last distribution is correlated with the jet multiplicity
distributions that we have already studied, we do not show it here.

\subsection{Gluino mass measurement}

Although measurement of sparticle masses can proceed with significant
precision at $e^+e^-$ colliders\cite{MUR} (given sufficient energy and
luminosity), the situation is expected to be much more difficult in
the environment of a hadron supercollider.  This is especially true of
measuring the mass of the gluino at the LHC.  Even in the ideal case
where all final state decay products of a gluino are tagged and
precisely measured, the invariant mass distribution created will be a
broad distribution between zero and $m_{\tg}-m_{\tz_1}$.  Real events
will contain overlapping decay debris from both of the sparticles
produced in the subprocess, in addition to significant amounts of QCD
radiation, and imperfect detector effects.

In the past, various methods have been proposed for sparticle mass
measurements at the LHC:
\begin{itemize}
\item In Ref.~\cite{assoc}, associated production processes such as
$\tg\tz_1$ were examined using parton level event generators. If these
events could be singled out, then the ambiguities from producing and
decaying {\it two} strongly interacting sparticles are by-passed. In
practice, rather hard cuts were required to separate the $\tg\tz_1$
events from $\tg\tg$ events.  Ultimately, it was concluded that this
reaction might be of use only if $m_{\tg}\alt 350$ GeV.
\item In Ref.~\cite{trilep}, it was shown that $m_{\tz_2}-m_{\tz_1}$
could be measured with significant precision from the end-point of the
dilepton invariant mass distribution.  However, the $\tw_1\tz_2\to
3\ell$ signal upon which this is based is only observable in a limited
region of parameter space.
\item In Ref.~\cite{bgh}, it was claimed that $m_{\tg}$ could be
measured to $15\%$ by focussing on same-sign isolated dilepton events
from $\tg\tg$ production, where each gluino decays via $\tg\to
q\bar{q}\tw_1$, with $\tw_1\to \ell\nu_{\ell}\tz_1$. However, these
calculations may be overly optimistic, since this study considered
only a single production mechanism ($\tg\tg$ production-- whereas same
sign dileptons can come from various SUSY sources), a single cascade
decay chain, and neglected effects of additional QCD radiation.
\end{itemize}

Here, we seek to measure the gluino mass in multi-jet$+\eslt$ events,
following a similar path to Ref.~\cite{bgh}. We proceed as follows.
First, we have a rough estimate of $m_{\tg}$ and $m_{\tq}$ by examining
signal to background levels versus cut parameter $E_T^c$. After
obtaining a relatively clean sample of signal events for an
appropriate $E_T^c$ value, we divide the event into two halves, using
eigenvectors from constructing the transverse sphericity $S_T$. We
then construct the invariant mass of jets in each of the two
hemispheres, using only jets with $p_T>E_T^c$.  Events are rejected if
there is only one qualifying jet in each hemisphere.  We then plot
as $M_{\rm est}$ the maximal value of the mass of each hemisphere.
Results are shown in Fig.~\ref{fig9} using $m_0=m_{1/2}$, $A_0=0$,
$\tan\beta =2$ and $\mu <0$. The corresponding value of $m_{\tg}$ is
shown in each frame.  Background level is shown by the shaded regions.
In these plots, as usual, all supersymmetric subprocesses are
contributing, with full cascade decays and QCD radiation effects, and
detector smearing as given in Sec.~II{\it a}.

We see in Fig.~\ref{fig9}{\it a}, using $E_T^c=100$ GeV for
$m_{\tg}\sim 300$ GeV, that the $M_{\rm est}$ distribution does indeed
have a large smear due to the various effects listed previously.
However, at least with our toy detector simulation, the two values of
$m_{\tg}$, which differ by $\sim 15\%$, appear distinguishable.  We
plot values of $M_{\rm est}$ for $m_{\tg}\sim 800$ GeV and
$E_T^c=350$~GeV in Fig.~\ref{fig9}{\it b}, and plot for $m_{\tg}\sim
1500$ GeV with $E_T^c=600$ GeV in Fig.~\ref{fig9}{\it c}.  Again,
values of $m_{\tg}$ differing by $\sim 15\%$ appear distinguishable.
We note, however, that using a too small value of $E_T^c$, or too
small a value of $p_T({\rm jet})$, can lead to a large amounts of
smearing in these distributions, and loss of distinguishability. These
distributions appear workable only for rather narrow ranges of cuts
that guarantee sufficient hemispheric separation of event debris, and
that only the hardest debris is used in constructing $M_{\rm est}$, so
that contamination of $M_{\rm est}$ is avoided.  In Fig.~\ref{fig10},
we show the same $M_{\rm est}$ plot, but now for $m_0=4m_{1/2}$, so
that $m_{\tq}\gg m_{\tg}$ (we show only the first two cases in
Fig.~\ref{fig9} since the last case is not observable above
background).  The distributions appear somewhat less distinguishable
than in Fig.~\ref{fig9}, so that a mass resolution of 15\% may be more
difficult to attain in this case; nevertheless, distributions from gluino
masses differing by 25\% appear to be distinguishable.
The jet multiplicity and/or
$B$-multiplicity (see next section) may have to be used to distinguish
whether $m_{\tq}\gg m_{\tg}$ or $m_{\tq}\sim m_{\tg}$.

\subsection{Detecting Higgs bosons via $B$-jets in cascade decay debris}

Multi-jet $+\eslt$ events from gluino and squark cascade decays are
expected to be unusually rich in heavy flavor content (mainly $B$
mesons, and $t$-quarks, if kinematically allowed) compared to SM
processes\cite{EARLY,BDKNT,BARTL}. This is due to a number
of effects.
\begin{itemize}
\item The $\tst_i$ and $\tb_i$ masses are driven to values lower than
the other squarks. This means decays like $\tg\to t\tst_i$ or $\tg\to
b\tb_i$ can be kinematically allowed, and dominate decay rates for
large enough values of $m_{\tg}$. Even if such decay modes are not
allowed, three-body $\tg$ decays to $t\bar{t}\tz_i$, $t\bar{b}\tw_i$
and $b\bar{b}\tz_i$ are enhanced relative to decays to other squarks
due to the larger propagator factor for the lighter top and bottom
squarks.
\item The various two- and three-body $\tg$ decays to third generation quarks
and squarks are further enhanced by the large third generation Yukawa
couplings\cite{BTWLOOP}.
\item $\tz_i\to\tz_j b\bar{b}$ decays are also enhanced by Higgs boson
mediated decays\cite{BDKNT}.
\item Higgs bosons can be produced in large quantities in $\tg$ and $\tq$
cascade decays. The various Higgs bosons frequently have dominant decays into
3rd generation particles, enhancing the heavy flavor content of $\tg$ and
$\tq$ events\cite{EARLY}.
\end{itemize}

In Table~III, we list the average $B$-meson multiplicity in
multi-jet$+\eslt$ events, for nine cases of SUGRA parameters. To
construct our tabulation, we have required $\eslt >200$ GeV,
$S_T>0.2$, and at least two jets with $p_T({\rm jet})>100$ GeV and
$|\eta ({\rm jet})|<2.5$. We then examine all jets with $p_T>50$ GeV
in the same rapidity interval.  If there is a $B$-hadron within
$\Delta R=0.5$ of the jet, it is tagged as a $B$ with 50\%
probability; otherwise, it is (mis)-tagged as a $B$ with 2\%
probability\cite{atlas}.  From Table~III, we see that in cases 1 and
2, for $m_0=100$ GeV, the $B$ multiplicity is rather low, $<n_B>\sim
0.9$, compared with cases 3--9, with higher values of $m_0$. This is
because {\it all} squarks and sleptons are relatively light, so any
enhancements in production and/or decay to third generation quarks are
small.  For a fixed value of $m_{\tg}$, the highest $B$ multiplicity
occurs for the cases with large $m_0$, which are dominated by $\tg\tg$
production, with enhanced cascade decays to heavy flavors:  here,
$<n_B>\sim 1.6$. Cases 3--5, with intermediate vaules of $m_0$, also
have intermediate values of $<n_B>\sim 1.1$. We also see that, as
discussed in Sec.~III{\it a}, the average jet multiplicity is smaller
for $m_0$ small, and larger for $m_0$ large.

Can one see direct evidence for Higgs production in cascade decay
events?  The light Higgs $H_{\ell}$ is usually produced somewhere down
the cascade decay chain via $\tz_2\to \tz_1 H_{\ell}$.  It has been
shown in Ref.~\cite{BBTW} that only in very limited regions of
parameter space is the decay $H_{\ell}\to\gamma\gamma$ likely to be
visible. We have investigated cases 1--9 of Table~II to see whether the
$H_{\ell}\to b\bar{b}$ decay can also be observed in cascade decay
events. We do this by constructing the invariant mass $m_{b\bar{b}}$
of all tagged $B$-jet pairs.  Results are shown in Fig.~\ref{fig11}.
First we show case 1 in Fig.~\ref{fig11}{\it a}, where $m_0=100$ GeV,
and $\tz_2$ dominantly decays to real sleptons and sneutrinos, so that
Higgs production in SUSY events is minimal. In this case, a Higgs
peak is not necessarily expected; we see a continuum distribution with only
a slight bump at $m_{H_\ell}$. In
Fig.~\ref{fig11}{\it b} (case 3), where both gluino and squark pair
production is important, $\tz_2\to\tz_1 H_{\ell}$ occurs at 94\%
branching ratio, the Higgs bump at $m_{b\bar b}=89$ GeV stands out
against the continuum from other tagged $B$-jet pairs in the SUSY
events. This is despite the fact that decays to $b$ and $t$ squarks
constitute 60\% of the gluino decays.  Higgs production from the
cascade decays of the first two generations of squarks remains
observable.  Finally, in Fig.~\ref{fig11}{\it c}, we show the
distribution for case 8, where $\tz_2\to\tz_1 H_{\ell}$ occurs with
98\% branching ratio.  However, in this case there are very many other
sources of tagged $B$-jets from $\tg$ cascade decays, especially
$\tg\to t\bar{t}\tz_{1,2}$, $b\bar{b}\tz_{1,2}$ and $\tg\to
t\bar{b}\tw_i$.  Moreover, since $\tg\tg$ production is by far the
dominant process, there is a very high $B$ multiplicity; then, the
Higgs boson is frequently produced along with other $B$-jets, so that
the combinatorial background tends to wash out the Higgs bump. The
Higgs signal was also detectable at the appropriate Higgs mass in case 4;
the signal did not stand out for the other four cases in Table~III.

In conclusion, it appears that while the $B$-multiplicity is
frequently enhanced above SM expectations, the $H_{\ell}\to b\bar{b}$
signal can be identified only in limited regions of the parameter
space: obviously, we need a substantial number of tagged $B$-jets from
Higgs decay, but also that these events should not (as, for instance,
in case 3) simultaneously contain other tagged $B$-jets which would
then produce a large combinatorial background. Another requirement for
the detection of the Higgs boson bump is that the rate from other
sources of tagged $B$-jets ({\it e.g.} $\tg \to \tst_1 t$ and $b\tb_1$
decays, where the squarks may be real or virtual) should not be
overwhelmingly large.

\section{Conclusions}

In this paper, we have performed a detailed analysis of the
multi-jet$+\eslt$ signal expected from production and decay of
supersymmetric particles within the framework of the minimal
supergravity model, in which sparticle masses and couplings are fixed
in terms of the parameter set (1).  Assuming an integrated luminosity
of 10 $\fb^{-1}$, we have shown that the $\eslt$ signal (mainly from
gluino and squark production and decay) should be observable in
experiments at the LHC for gluino masses ranging from values
accessible to Fermilab Tevatron experiments, up to nearly $m_{\tg}\sim
1300$ GeV ($m_{\tg}\sim 2000$ GeV ) for $m_{\tq}\gg m_{\tg}$ (for
$m_{\tq}\sim m_{\tg}$), confirming earlier studies by the ATLAS
collaboration\cite{atlas}. We expect that the reach (in terms of
sparticle masses) is not very sensitive to the details of the model as
long as R-parity is conserved.  Comparing the LHC reach in terms of
$m_{\tg}$ and $m_{\tq}$ with somewhat subjective upper bounds from
fine tuning constraints\cite{finetun} (recall that the resolution of
the fine-tuning problem provided the original motivation for
phenomenological SUSY), it seems likely that LHC can perform a
thorough search for $R$-conserving weak scale supersymmetry.

We show the SUSY reach via the multi-jets$+\eslt$ channel in the $m_0\
vs.\ m_{1/2}$ plane of the minimal SUGRA model with gauge coupling
unification and radiative electroweak symmetry breaking in
Fig.~\ref{fig5}--\ref{fig7}.  The relatively small parameter set
yields a complete, highly correlated sparticle spectrum, so that the
plot of the reach in the $\eslt$ channel can be compared with previous
studies on the reach for sleptons and charginos/neutralinos. By
comparing these various regions in Figs.~\ref{fig5}--\ref{fig7}, we
see that of the channels studied, the reach in multi-jets$+\eslt$ is
by far the largest. However, it should be kept in mind that
multilepton signals ({\it e.g.} same-sign dileptons and trileptons)
from cascade decays of gluinos and squarks may also probe some or all
of this region. These signals would be especially important if $m_0$
is very small, or if gluino decays to third generations fermions and
sfermions are strongly enhanced.  These multilepton signals could then
provide a striking confirmation of a supersymmetric signal discovery
in the multi-jet$+\eslt$ channel.

We have also studied what further information about sparticle
properties can be obtained by a careful study of the $\eslt$ sample.
We have used ISAJET to demonstrate that jet multiplicity can be a
useful tool for detecting the presence of squarks in the $\eslt$
sample, and indirectly inferring the squark to gluino mass ratio.  Of
course, the distribution of jet multiplicity is sensitive to the
sparticle mass, so that this method is useful only after an estimate
of the mass is obtained.  We have further demonstrated that it might
be possible to obtain a measure of the gluino mass by dividing each
event in two halves, and using the greater of the masses in the two
hemispheres as an estimator of $m_{\tg}$. We see from the
distributions in Figs.~\ref{fig9} and \ref{fig10} that gluino masses
differing by 15--25\% might be possible to distinguish. Since we have
included the production of all sparticles in our simulation as well as
contamination to the hemispheres from QCD jets, we believe that it
would be worth testing this technique to see if it survives a detailed
detector simulation. Finally, we have studied if signals from Higgs
bosons produced by cascade decays of gluinos and squarks and decaying
via $H_{\ell}\to b\bar{b}$ are detectable. We conclude that, with
reasonable $B$ tagging efficiencies, this is possible but only for
favourable ranges of parameters, where events containing Higgs bosons
are relatively free of other $B$-jets, and further, that $B$-jets from
other SUSY sources do not overwhelm the Higgs signal.  It should
nonetheless be kept in mind that SUSY events frequently tend to be
rich in central $B$-jets, so that this may provide another handle for
distinguishing SUSY from the SM, and gaining information on the squark
to gluino mass ratio.

%
\acknowledgments

This research was supported in part by the U.~S. Department of Energy
under contract number DE-FG05-87ER40319, DE-AC02-76CH00016, and
DE-FG-03-94ER40833.
%
%
%
%

\newpage
%
%

\iftightenlines\else\newpage\fi

\begin{table}
\caption[]{Six parameter choices used for cross section evaluation of
multi-jet $+\eslt$ events.
We take $A_0=0$, $\tan\beta =2$ and $\sgn(\mu )<0$. Also, $m_t =170$ GeV.}

\bigskip

\def\d{\phantom{0}}

\begin{tabular}{lcccc}
$Case$ & $m_0$ & $m_{1/2}$ & $m_{\tg}$ & $m_{\tq}$ \\
\tableline
1 & \d100 & 100 & \d290 & \d270 \\
2 & \d400 & 100 & \d310 & \d460 \\
3 & \d300 & 300 & \d770 & \d720 \\
4 & 1200 & 300 & \d830 & 1350 \\
5 & \d600 & 600 & 1400 & 1300 \\
6 & 2000 & 500 & 1300 & 2200 \\
\end{tabular}
\end{table}
\begin{table}
\caption[]{Details of background calculation after cuts using cut
parameter $E_T^c=500$ GeV. We list the background process ($BG$), the range of
hard scattering $p_T$ ($p_T^{HS}$) over which they were evaluated,
the number of events generated ($N$),
number of events passing cuts ($N_{cut}$), total BG
cross section ($\sigma_{tot}$ in $\fb$), and background cross section after
cuts
($\sigma_{cut}$ in $\fb$). We compute the combined background to be
$\sigma =1.86\pm 0.36\ \fb$, which yields a 95\% CL upper limit of
$\sigma =2.44\ \fb$, which we use for the computation of the LHC reach.
We take $m_t =170$ GeV.}

\bigskip

\def\d{\phantom{0}}

\begin{tabular}{llcccc}
$BG$ & $p_T^{HS}$ & $N$ & $N_{cut}$ & $\sigma_{tot}$ & $\sigma_{cut}$ \\
\tableline
$QCD$ & 200-1000 & $2.843\times 10^7$ & 0 & $8.9\times 10^7\ \fb$ & --- \\
$QCD$ & 500-1000 & $10^5$ &0 & 1085 & --- \\
$QCD$ & 800-2000 & $2\times 10^6$ & 17 & $8.0\times 10^5$ &
$(6.8\pm 1.6)\times 10^{-1}$ \\

$QCD$ & 2000-3000 & $2.5\times 10^5$ & 21 & 103 &
$(8.7\pm 1.9)\times 10^{-3}$ \\

$t\bar t$ & 500-1000 & $5.5\times 10^5$ & 0 & $7.2\times 10^5$ & --- \\

$t\bar t$ & 1000-2000 & $1\times 10^5$ & 14 & $0.48$ &
$(6.7\pm 1.8)\times 10^{-3}$ \\

$W+$jets & 10-800 & $1.78\times 10^7$ & 0 & $3.1\times 10^7$ & --- \\

$W+$jets & 300-800 & $4\times 10^5$ & 9 & $9.9\times 10^3$ &
$(2.2\pm 0.74)\times 10^{-1}$ \\

$W+$jets & 800-2000 & $8\times 10^5$ & 1105 & $0.89$ &
$(1.2\pm 0.04)\times 10^{-1}$ \\

$Z+$jets & 100-300 & $5\times 10^4$ & 0 & $147$ &
--- \\

$Z+$jets & 300-1000 & $2.5\times 10^4$ & 6 & $3\times 10^3$ &
$(7.3\pm 3.0)\times 10^{-1}$ \\

$Z+$jets & 1000-2000 & $1\times 10^5$ & 1314 & 7.0 &
$(9.1\pm 0.25)\times 10^{-2}$ \\

\end{tabular}
\end{table}

\begin{table}
\caption[]{Nine cases examined for $B$-multiplicity, and the possibility of
reconstructing a Higgs mass via $m(b\bar{b})$. We take $A_0=0$, and
$\mu <0$, except for case 4, for which $\mu >0$. All mass quantities are
in units of GeV.}

\bigskip

\def\d{\phantom{0}}

\begin{tabular}{lcccccccc}
Case & $m_0$ & $m_{1/2}$ & $\tan\beta$ & $m_{H_{\ell}}$ &
$\langle n(b\hbox{-tag})\rangle$ & $\langle p_T(b\hbox{-tag})\rangle$ &
$\langle n(\hbox{no-tag})\rangle$ &
$\langle p_T(\hbox{no-tag})\rangle$ \\
\tableline
1 & \d100 & 300 & \d2  & \d87.7  & 0.90 & 127 & 4.5 & 124 \\
2 & \d100 & 300 & 10 & 116.6 & 0.90 & 126 & 4.5 & 124 \\
3 & \d300 & 300 & \d2 & \d88.4 & 1.16 & 121 & 4.6 & 121 \\
4 & \d300 & 300 & \d2 & 102.5 & 1.18 & 124 & 4.7 & 121 \\
5 & \d300 & 300 & 10 & 116.6 & 1.04 & 123 & 4.7 & 121 \\
6 & \d600 & 300 & \d2 & \d90.0 & 1.65 & 118 & 5.1 & 113 \\
7 & \d600 & 300 & 10 & 117.2 & 1.31 & 124 & 5.7 & 119 \\
8 & 1000 & 275 & \d2 & \d92.3 & 1.59 & 126 & 5.8 & 123 \\
9 & 1000 & 350 & 10 & 120.5 & 1.68 & 128 & 6.2 & 122 \\
\end{tabular}
\end{table}



\begin{figure}
\caption[]{Total cross section for strongly interacting sparticle pairs
(solid),
associated production of gluinos or squarks with charginos or neutralinos
(dot-dashed) and chargino/neutralino production (dashes),
for {\it a}) $m_{\tq}=m_{\tg}$ and {\it b}) $m_{\tq}=2m_{\tg}$, as a
function of $m_{\tg}$, for $pp$ collisions at $\sqrt{s}=14$ TeV.}
\label{fig1}
\end{figure}

\begin{figure}
\caption[]{Cross section in $\fb$ after cuts,
as a function of cut parameter $E_T^c$, for total background
($\times$'s), and for the six signal cases listed in Table 1.}
\label{fig2}
\end{figure}

\begin{figure}
\caption[]{{\it a}) Cross section in $\fb$ after cuts,
as a function of $A_0/m_0$, for parameter choices shown. In {\it b}),
we show variation in the cross section for Table 3 case 3 when $\tan\beta$
is varied.}
\label{fig3}
\end{figure}

\begin{figure}
\caption[]{Background cross sections vs.\ $E_T^c$ obtained from
various samples of events. The $p_T^{HS}$ range of the hard scattering
subprocesses is listed.}
\label{figbkg}
\end{figure}

\begin{figure}
\caption[]{Signal cross sections for $m_0=m_{1/2}$ (squares) and
$m_0=4m_{1/2}$ (triangles), after cuts, using $E_T^c=500$ GeV, as a function
of $m_{\tg}$. We also show the $5\sigma$ background cross section
assuming an integrated luminosity of $10$ $\fb^{-1}$. We take
$A_0=0$, $\tan\beta =2$, and $\mu <0$.}
\label{fig4}
\end{figure}

\begin{figure}
\caption[]{In {\it a}), we show regions of SUGRA parameter space
excluded by theory (bricks), excluded by experiments (slashed), and
regions explorable by LHC assuming integrated luminosity of $10\ \fb^{-1}$,
for $A_0=0$, $\tan\beta =2$, and $\mu <0$. We also show regions
explorable via slepton searches at LHC, and Higgs and chargino searches
at LEP II. Much of the region below the spoiler mode ($\tz_2\to\tz_1 Z$ and
$\tz_2\to\tz_1 H_l$) contour is explorable via isolated trileptons at both
Tevatron and LHC. In {\it b}), we show the same plane, but with various
mass contours for $m_{\tg}$ and $m_{\tq}$ (averaged over 1st generation).}
\label{fig5}
\end{figure}

\begin{figure}
\caption[]{Same as Figure 5, except for $\mu >0$.}
\label{fig6}
\end{figure}

\begin{figure}
\caption[]{Same as Figure 5, except for $\tan\beta =10$.}
\label{fig7}
\end{figure}

\begin{figure}
\caption[]{Fractional jet multiplicity after cuts (signal plus background),
for six cases listed in Table 1, for different values of $E_T^c$. The jet
$p_T$ cut has been relaxed to $p_T({\rm jet})>50$ GeV.}
\label{fig8}
\end{figure}

\begin{figure}
\caption[]{Distribution in $M_{est}$ for different values of $m_{\tg}$
and cut parameter $E_T^c$, along with background contribution (shaded).
$M_{est}$ is formed after cuts by dividing the event into hemispheres
via the transverse sphericity eigenvector, forming the invariant mass of
the jets, and taking the maximal value. Events with only one jet in a
hemisphere are rejected. We take $m_0=m_{1/2}$.}
\label{fig9}
\end{figure}

\begin{figure}
\caption[]{Same as Fig.~10, except we take $m_0=4m_{1/2}$.}
\label{fig10}
\end{figure}

\begin{figure}
\caption[]{Mass distributions for pairs of tagged $b$ jets.
(a) Case 1 from Table~III. Higgs production is small.
(b) Case 3 from Table~III. $B(\tz_2\to\tz_1 H_{\ell}) = 94\%$, and peak
is observed at $m_{H_\ell}$.
(c) Case 8 from Table~III. $B(\tz_2\to\tz_1 H_{\ell}) = 98\%$, but more
combinatorial background.}
\label{fig11}
\end{figure}

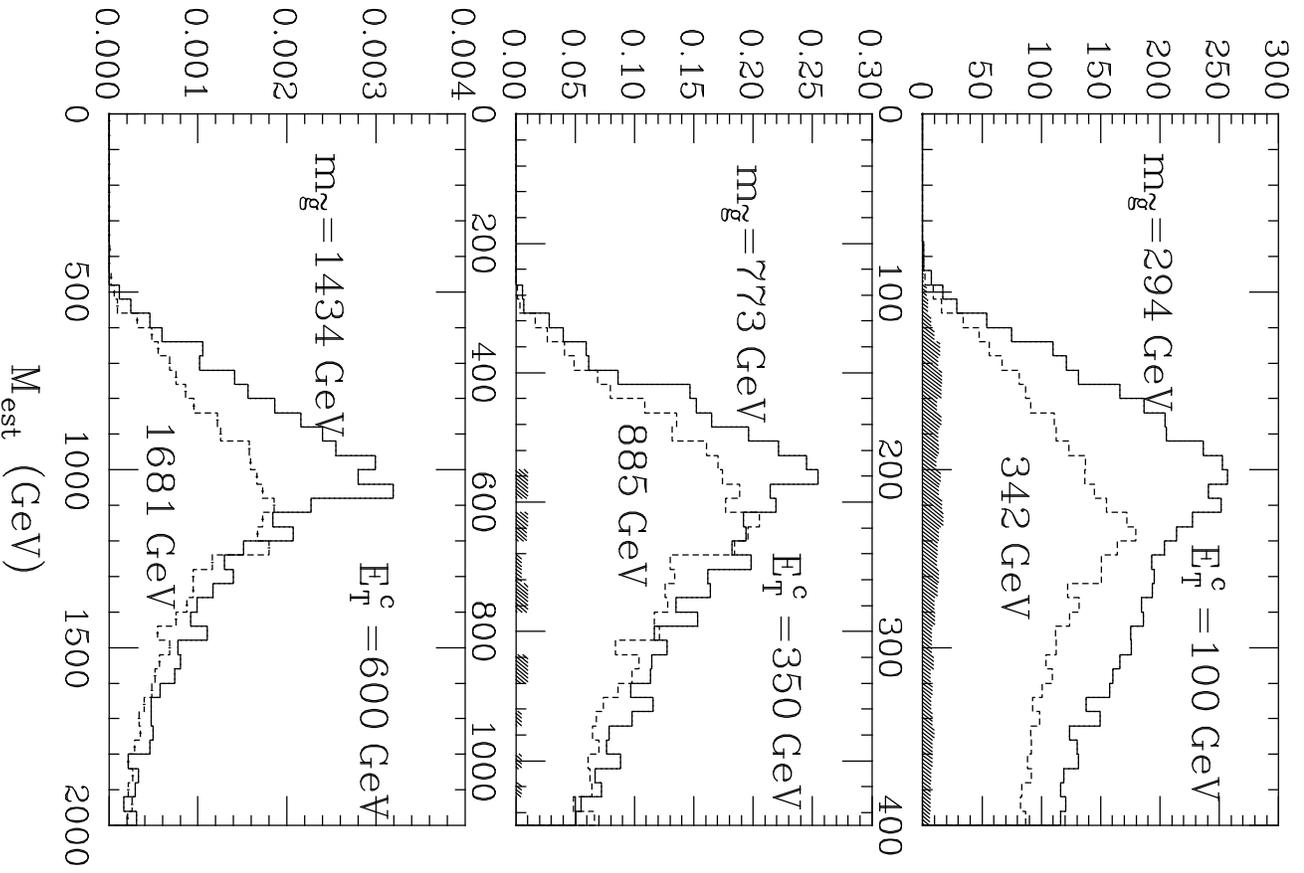
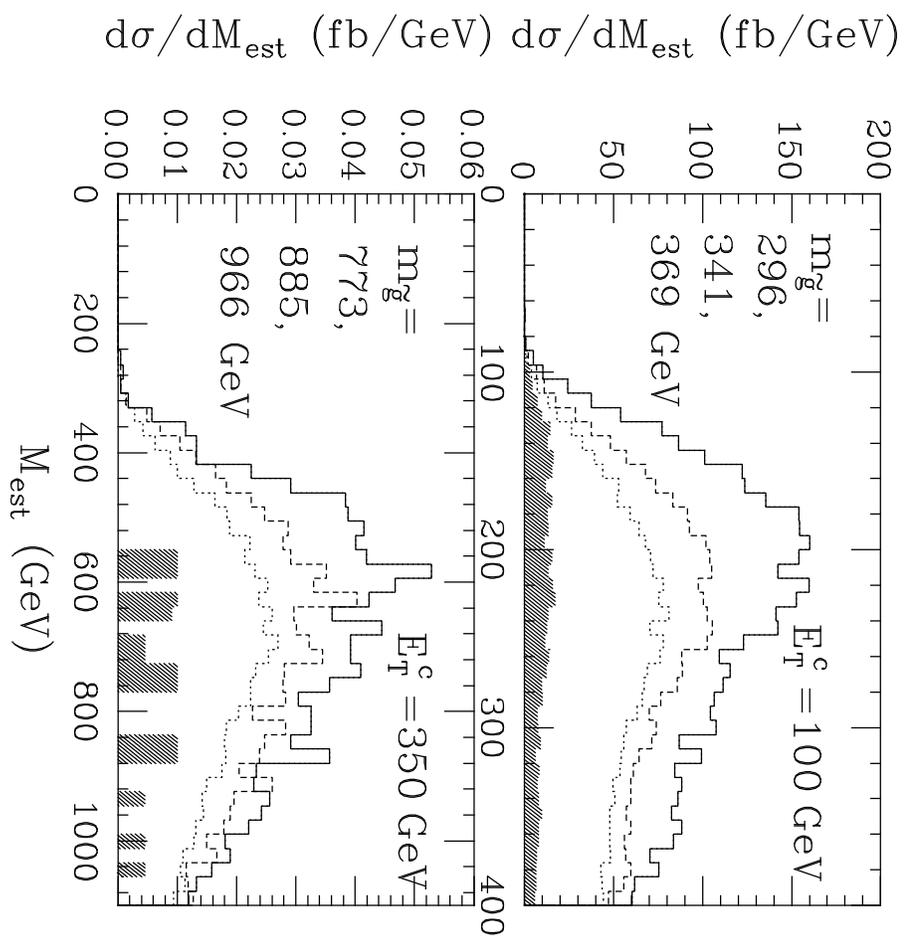
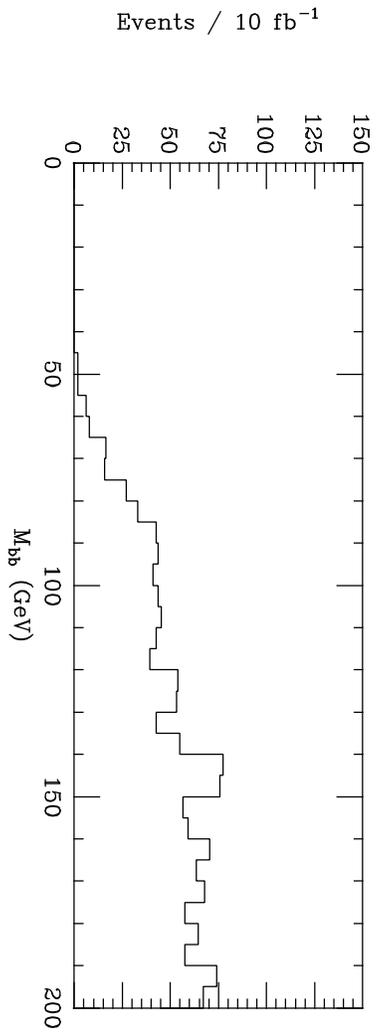
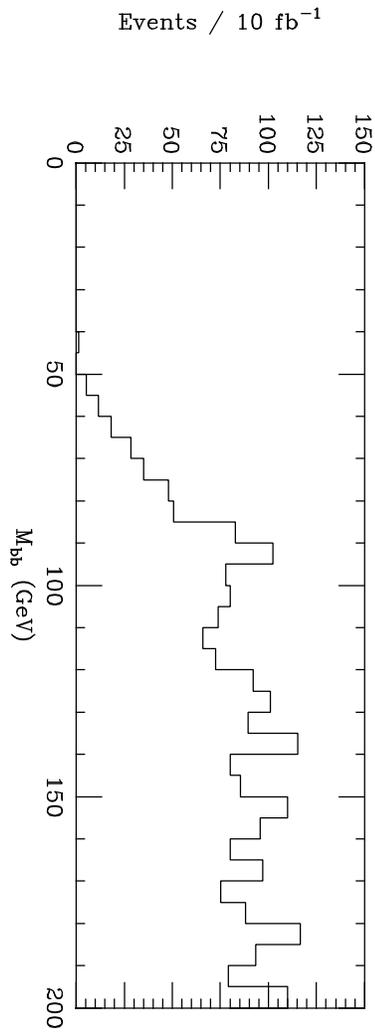
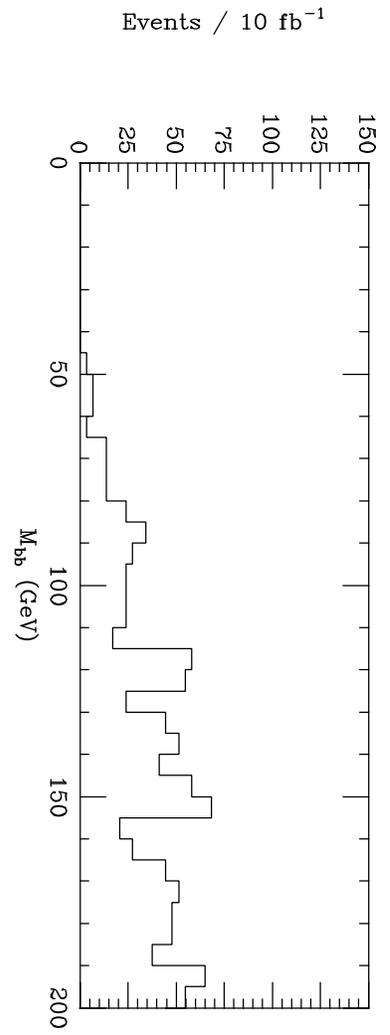
\end{document}